\documentstyle[11pt,epsfig]{article}

\def\laq{~\raise 0.4ex\hbox{$<$}\kern -0.8em\lower 0.62 
ex\hbox{$\sim$}~}
\def\gaq{~\raise 0.4ex\hbox{$>$}\kern -0.7em\lower 0.62 
ex\hbox{$\sim$}~}
\def\vk{\vec{k}}
\def\vp{\vec{p}}
\def\vx{\vec{x}}
\def\vy{\vec{y}}

\begin{document}
\begin{titlepage}

\begin{flushright}
CERN-TH/2003-041 
\end{flushright}

\vspace*{1cm}

\begin{center}
\Large{\bf  Cosmological Perturbations from \\a New-Physics Hypersurface}

\vspace*{1cm}

\large{ V. Bozza${}^{1,2,3}$,
M. Giovannini${}^{4}$ and G. Veneziano${}^{4}$}
\bigskip
\normalsize

{\sl $^{(1)}$  Dipartimento di Fisica
``E. R. Caianiello", Universit\`a di
Salerno, \\
Via S. Allende, 84081 Baronissi (SA), Italy\\
\vspace{0.3cm}
$^{(2)}$ INFN, Sezione di
Napoli, Gruppo Collegato di Salerno, Salerno, Italy} \\
\vspace{0.3cm}
{\sl $^{(3)}$ Enrico Fermi Center, via Panisperna 89/A, 00184 Roma, Italy}\\
\vspace{0.3cm}
{\sl $^{(4)}$  Theoretical Physics Division, CERN,
CH-1211 Geneva 23, Switzerland}

\vspace*{5mm}

\begin{abstract}
 Within a broad  class of inflationary 
models we critically analyse  the way initial quantum
fluctuations on a new-physics hypersurface (NPH) affect
standard predictions for
large-scale cosmological perturbations. We find that
these so-called transplanckian effects crucially depend  
on the definition of the  ``vacuum state" in particular
on which Hamiltonian is minimized on the NPH in order to select such a state. 
Transplanckian effects can be made much smaller than
previously suggested if sufficiently ``adiabatic" Hamiltonians are minimized.
\end{abstract}
\end{center}
\end{titlepage}

There is general consensus that presently observed CMB anisotropies, as
well as the large-scale structure of our visible Universe, originate
 from the parametric amplification of quantum-mechanical fluctuations occurring
during inflation (for a review, see e.g. \cite{ks}). Present large-scale
 experiments thus offer a window on
physics at the energy scales at which inflation took place. Since the
experimental value of CMB temperature fluctuations bounds the 
Hubble parameter during inflation by $10^{-6}-10^{-5} M_{\rm P}$, this implies that present
large-scale observations are  sensitive to physics at energy scales
that can be, at most, as large as  $10^{-3} M_{\rm P} \sim M_{\rm GUT}$. While this is certainly
 a very interesting energy scale for particle physics, it is definitely
below the scale of quantum gravity or, presumably, the scale
characterizing new physics in  any realistic string theory.

There is instead much debate \cite{kem,gree,bra,kal1,star} about the possibility that present large-scale
experiments may test physics at even shorter scales, down to the
string/Planck-length scales or even below. A particularly important issue is that of
finding out the order of these ``transplanckian" corrections in the (supposedly small) parameter
$H/\Lambda$, where $H$ is a typical Hubble parameter during inflation and $\Lambda$ is the scale of
some unknown new physics. Although large effects can result from assuming that the new physics
breaks some well-established low-energy principles, one would like to establish the existence and 
order of magnitude
of the effects in a more model-independent way.

An interesting step in this direction goes as follows \cite{dan} (see also \cite{GV}).
Any scale of physical interest today was very tiny if 
blue-shifted  back till the beginning of inflation. In fact,
 unless the duration
of inflation is  just the minimal one for solving 
the standard cosmological problems,
even the largest relevant cosmological scales today were very much sub-Planckian
at the beginning of inflation.
To reinforce this statement, let us remark 
that the difference between initial classical and quantum fluctuations
is  that the former, by definition, are at super-Planckian length scales,
and are therefore much beyond our horizon --hence unobservable-- today,
while the latter must have also been  present at sub-Planckian scales. This
characterization of the two kinds of initial perturbation 
is crucial if one wants
to argue that inflation washes out any initial (and arbitrary) classical inhomogeneity
while replacing it by a  spectrum of fluctuations that is fully normalized
by the uncertainty principle of quantum mechanics.

The problem with the above picture is that it forces one to normalize
quantum fluctuation in a regime (the transplanckian one) where, in general, physics is
not known. Furthermore, if the standard formula is used, say for tensor metric 
perturbations,
\begin{equation}
\delta_h (k) = k^{3/2} h(k) \sim \ell_{\rm P} k/a,
\label{tensor}
\end{equation}
we see that the fluctuations of the metric become $O(1)$ as soon as
$\omega \equiv k/a = M_{\rm P}$ \footnote{Note that this is not the case for tensor perturbations
 in pre-big bang cosmology \cite{PBB}, since the relevant scale factor
 is the one in the Einstein frame and blows
up, rather than vanishes, at $t \rightarrow -\infty$.}.

In order to avoid the above difficulty, while preserving the crucial
distinction between classical and quantum fluctuations, it can be argued, instead \cite{GV},
  that, while classical fluctuations are  given for all
scales in excess of some cut-off length $l = \Lambda^{-1}$
 on one and the same initial space-like
hypersurface, quantum fluctuations are created at all times, and should only
be looked at, for each physical scale $\omega$, after an ``initial" 
($k$-dependent) time defined by $\omega(t) = k/a(t) = \Lambda$. In other words, 
in this new way of looking at initial conditions,
both classical and quantum fluctuations are only considered above the scale
$\Lambda^{-1}$, but, while the former are given on a conventional space-like
hypersurface, the latter are given \cite{dan,GV}  on the unusual ``new-physics hypersurface"
(NPH) defined by $\omega(t) = \Lambda$, hence in a way that mixes 
coordinates and momenta. 

If this point of view is taken the question arises of the possible effects
of the NPH on present observations. It has been argued, for instance, 
that in the de Sitter case 
 corrections in the two-point function of the metric fluctuations are linear 
in the ratio $H/\Lambda$  \cite{dan}, computable, and  possibly observable \cite{dan2} (see also \cite{cl,cl2,cl4} for  different 
perspectives). 
While some aspects of this suggestion may be theoretically justified
 according to our previous discussion, 
 the mere existence of a cut-off does not allow, in our opinion, 
to compute accurately any specific correction to observables unless 
the nature of the physics above the cut-off is, even roughly, specified.
Suppose, for instance, that, for some reason, we take the
 initial state of the fluctuations on the NPH to be the one obtained from 
 the unitary evolution of
the state minimizing the Hamiltonian at $t \to -\infty$ \footnote{Such an assumption may not be
 as crazy as it looks, particularly when, as $t \to -\infty$,  space-time becomes
trivial. There are presumably no transplanckian effects due to just having a single particle/string
carry momentum in excess of $M_{\rm P}$, as is well known for
the vertex operators of string theory in flat space-time.}. In this case the two-point function 
will not have any correction related to the ``initial'' time $t_0(k)$ lying on the NPH.

Assume instead, as  in \cite{dan,dan2}, that we do not know 
anything of the physics above $\Lambda$ or, if we prefer, about what happened to the past of the NPH. 
Logic would suggest that, in this case, very little, if anything, can be said  
about the corrections to the standard predictions. The simple technical reason for 
this  is that, in order to compute correlation functions at some late time, 
we have to know the initial state on the NPH and this is all but impossible without
some knowledge of the physics above the cut-off. 

In this Letter we shall explore the sensitivity of these
corrections to the choice of the initial state on the NPH. 
In the absence of a specific knowledge of ultra-high energy physics, we shall demand 
that  such initial state  minimizes the Hamiltonian on the NPH. If this were a unique prescription,
such an assumption would lead to definite predictions on the form and the magnitude
of the corrections. 
Unfortunately, in 
time-dependent problems, there is no conserved Hamiltonian and the possibility 
of performing time-dependent field-redefinitions  leads  to different 
Hamiltonians  connected among themselves by simple canonical transformations. While the 
(classical or quantum mechanical) evolution is indeed independent of the choice of the Hamiltonian, 
the minimization of one or the other on the NPH leads to  different initial states and, consequently, 
to different corrections in the power spectrum. As we shall see,  in the argument 
of \cite{dan,dan2} there is an implicit  assumption that ``transplanckian'' effects 
lead to the minimization of a particular Hamiltonian, the one that turns out to give ``large" corrections.
After a general discussion of the quantum mechanical evolution of 
the system we shall look at the consequences of minimizing various Hamiltonians characterized by their
``adiabaticity" i.e. by the way they go to the free Hamiltonian as $t \to -\infty$. 
Amusingly, we  will find a strict correlation between the degree of ``adiabaticity" and 
the size of the corrections, suggesting that the latter should be much smaller than claimed in 
\cite{dan,dan2} and possibly even smaller than argued by other authors \cite{kal}.

Consider, for simplicity, a spatially flat Friedmann-Robertson-Walker
geometry whose line element can be written, in conformal time, as 
\begin{equation}
ds^2 = a^2(\eta) [ d\eta^2 - d\vec{x}^2],
\label{met}
\end{equation}
where $a(\eta)$ is the scale factor whose rate of variation $a'/a$ will be 
denoted by ${\cal H}$ (we shall use a prime to denote derivatives with
respect to conformal time).

Consider now, in the background (\ref{met}), 
 some metric fluctuations, for instance the tensor perturbations
of the geometry having the nice feature of being decoupled from 
the fluctuations of the matter sources. The time evolution of 
each of the two tensor polarization will  be obtained by perturbing to second order 
the gravity action, taken to be, for simplicity, 
of the Einstein-Hilbert form. The dynamics of each tensor polarization $h$
will then be described by \cite{gr}
\begin{equation}
S = \frac{1}{ 4 \ell_{\rm P}^2}\int d^4 x ~a^2~ \eta^{\mu\nu} \partial_{\mu} h
\partial_{\nu} h ,
\label{ac1} 
\end{equation}
where $\ell_{\rm P} = M_{\rm P}^{-1}= \sqrt{8 \pi G}$ and $\eta_{\mu\nu}$ is the 
Minkowski metric.

Let us assume, following \cite{dan,GV}, that the  effective description 
of the tensor modes of the geometry provided by Eq. (\ref{ac1}) 
 is only valid up to some 
typical scale $\Lambda$ at which the laws of some new/unknown 
physics replace Eq. (\ref{ac1}). This assumption 
 amounts to demanding  a cut-off on the physical 
momenta of the gravitons, i.e. 
\begin{equation}
k/a(\eta) \leq \Lambda,
\label{cond}
\end{equation}
 where $k$ is the comoving wavenumber.
In any kind of inflationary background, Eq. (\ref{cond}) 
is saturated at some time $\eta_{0}(k)$ defining the NPH on which    
initial conditions for the fluctuations are given.

As discussed above, owing to the time dependence appearing in the action (\ref{ac1}), 
different Hamiltonians can be used to describe the evolution
of the system. Defining, for the sake of simplicity, 
\begin{equation}
\Psi = \frac{  h}{ \sqrt{2}~\ell_{\rm P}},
\label{Psi}
\end{equation}
a  possible selection of  different Hamiltonians can be the following 
\begin{eqnarray}
&& H^{(1)}(\eta) = \int d^3 x \frac{1}{2}\biggl[ \frac{ \Pi^2}{ a^2 } + a^2 (\partial_{i} \Psi)^2\biggr],
\label{H1}\\
&& H^{(2)}(\eta) = \int d^{3} x \frac{1}{2}\biggl[ \pi^2 +  2{\cal H} \psi \pi + (\partial_{i} \psi)^2 
\biggr],
\label{H2}\\
&& H^{(3)}(\eta) = \int d^{3} x \frac{1}{2}\biggl[ \tilde{\pi}^2 + 
(\partial_{i} \psi)^2 -  \frac{a''}{a} \psi^2\biggr],
\label{H3}
\end{eqnarray}
where 
\begin{equation}
\Pi = a^2 \Psi',~~~~\psi = a \Psi, ~~~~~\pi = \frac{\Pi}{a},~~~~~\tilde{\pi} = \psi',
\label{def}
\end{equation}
are the relations between the different canonical fields. It is not difficult to check that 
it is possible to go from one Hamiltonian to the other through a
suitable canonical transformation. For instance, the transformation 
$H^{(2)}(\eta) \to H^{(3)}(\eta)$ is generated, in a standard way, by 
\begin{equation}
{\cal F}_{2\to 3} ( \psi, \tilde{\pi}, \eta) = \int d^{3} x \biggl( \psi \tilde{\pi}
 - \frac{{\cal H}}{2} \psi^2\biggr),
\label{2to3}
\end{equation}
 a functional of the old fields and of the new momenta $\tilde{\pi}$.
By differentiating the generating functional, we obtain 
the relation between the old momenta (i.e. $\pi$) and the new ones, 
as well as a change in the Hamiltonian
\begin{eqnarray} 
&& \pi = \tilde{\pi} - {\cal H} \psi,
\label{pigen1}\\
&& H^{(2)}( \psi, \pi, \eta) \to H^{(3)}(\psi,\tilde{\pi},\eta) = H^{(2)}(\psi, \pi,\eta) + \frac{\partial{\cal F}_{2\to3}}{\partial\eta}.
\label{hamgen1}
\end{eqnarray}
Bearing in mind Eqs. (\ref{2to3}) and (\ref{pigen1}), the right-hand side of 
Eq. (\ref{hamgen1}) leads exactly to Eq. (\ref{H3}). With similar considerations,
all the Hamiltonians (\ref{H1})--(\ref{H3}) can be related to one another 
by suitable canonical transformations.

Fully equivalent classical evolutions should be expected  by solving the 
appropriate Hamilton equations with the Hamiltonians
(\ref{H1})--(\ref{H3}). This statement is also true at the quantum level (see e.g. \cite{mat}) because 
the different classical actions corresponding to Eqs. (\ref{H1})--(\ref{H3}) 
only differ by total derivatives. Hence, thinking for instance in terms of a functional integral approach,  
the transitions amplitudes differ, at most, by a field-dependent phase since the total 
derivatives appearing in the classical action, once inserted in the 
path integral, can be explicitly integrated.
 
As an  example, consider the case of  $H^{(2)}(\eta)$ whose associated
Lagrangian density is 
\begin{equation}
{\cal L}^{(2)}(\vec{x},\eta) = \frac{1}{2} \biggl[ {\psi'}^2 - 2 {\cal H} \psi\psi' + 
{\cal H}^2 \psi^2 - (\partial_{i} \psi)^2\biggr].
\label{L1}
\end{equation}
Comparing Eq. (\ref{L1})  to the Lagrangian density associated with  $H^{(3)})(\eta)$, we can notice 
that they differ by a total derivative
\begin{equation}
{\cal L}^{(3)}(\vec{x},\eta) = {\cal L}^{(2)}(\vec{x},\eta) + \frac{d {\cal D}}{d\eta}, 
~~~~~{\cal D} =  \frac{ {\cal H}}{2} \psi^2.
\label{totder}
\end{equation}
The wave-functional computed with (\ref{totder}) only 
 differs from the one
computed with (\ref{L1}) by a field-dependent phase, namely
\begin{equation}
\Phi^{(3)}[\psi] = e^{i {\cal D}(\psi,\eta)} \Phi^{(2)}[\psi].
\label{PH}
\end{equation}

The expectation value of a generic observable 
${\cal O}[\hat{\psi},\hat{\pi}]$ can be computed  using either
$\Phi^{(2)}[\psi]$ or $\Phi^{(3)}[\psi]$.  However, 
 if we compute such an expectation 
value using $\Phi^{(3)}[\psi]$, we have to bear in mind that the canonical momentum
($\hat{\pi}$) acts non-trivially on it,   its 
action being fully specified by the transformation (\ref{pigen1}). The conclusion  is that
the expectation value of any operator ${\cal O}[\hat{\psi},\hat{\pi}]$ is independent
 on the Hamiltonian one is using (we shall see below an example of this fact in the context of
cosmological perturbations).
In spite of the above equivalence, 
the state minimizing one of the Hamiltonians  at the initial time $\eta_0$ 
does depend on which one of  (\ref{H1})--(\ref{H3}) is chosen. 
Before discussing the effect of choosing different initial states, let us discuss
 the Heisenberg-picture evolution of the operators 
that holds for any choice of the Hamiltonian.

Working in Fourier space,
\begin{equation}
\hat{\psi}(\vec{x},\eta) = \frac{1}{(2\pi)^{3/2} } \int d^3 k  ~\hat{\psi}_{\vk}(\eta) e^{- i \vec{k} \cdot \vec{x} },~~~~~~~~
 \hat{\pi}(\vec{x},\eta) = \frac{1}{ (2\pi)^{3/2} } \int d^3 k  ~\hat{\pi}_{\vk}(\eta) e^{- i \vec{k} \cdot \vec{x} },
\label{expansion}
\end{equation}
with $\hat{\psi}_{\vk} = \hat{\psi}_{-\vk}^{\dagger},  
\hat{\pi}_{\vk} = \hat{\pi}_{-\vk}^{\dagger}$, 
the evolution in the Heisenberg picture reads
\begin{equation}
\left(\matrix{ \hat{\psi}_{\vk}(\eta)\cr
                \hat{\pi}_{\vk}(\eta)\cr}\right) 
= \left(\matrix{A_{\vk}(\eta,\eta_{0})& B_{\vk}(\eta,\eta_0)\cr
 C_{\vk}(\eta,\eta_{0})& D_{\vk}(\eta,\eta_0)\cr}\right)\left(\matrix{ \hat{\psi}_{\vk}(\eta_{0})\cr
                \hat{\pi}_{\vk}(\eta_0)\cr}\right),
\label{trans}
\end{equation}
where 
\begin{eqnarray}
A_{\vk}(\eta,\eta_0)&=&i\biggl[ g_{k}(\eta_0) f^{\star}_{k}(\eta)
 - g^{\star}_{k}(\eta_0) f_{k}(\eta) \biggr],~
B_{\vk}(\eta,\eta_0) = i\biggl[ f_{k}(\eta) f^{\star}_{k}(\eta_0)
 - f^{\star}_{k}(\eta)  f_{k}(\eta_0) \biggr],
\nonumber\\
C_{\vk}(\eta,\eta_0)&=&i\biggl[ g_{k}(\eta_0) g^{\star}_{k}(\eta)
 - g^{\star}_{k}(\eta_0) g_{k}(\eta) \biggr],~
D_{\vk}(\eta,\eta_0) = i\biggl[ g_{k}(\eta) f^{\star}_{k}(\eta_0)
 - g^{\star}_{k}(\eta) f_{k}(\eta_0) \biggr].
\label{abcd}
\end{eqnarray}
In Eqs. (\ref{abcd}) $f_{k}(\eta)$ and $g_{k}(\eta)$ denote the mode functions that are, respectively, solutions of 
the Heisenberg evolution equations for a (generic) pair ($\hat{\psi}$, $\hat{\pi}$) of canonically conjugated operators.
At every time, for consistency with the canonical commutation relations, the phases and amplitudes 
of the mode functions are subjected to the Wronskian condition
\begin{equation}
f_{k}(\eta) g^{\star}_{k}(\eta) - f^{\star}_{k}(\eta) g_{k}(\eta) = i.
\label{wr1}
\end{equation}
In Eqs. (\ref{abcd}), with the condition (\ref{wr1}), for $\eta\to \eta_{0}$,
$C_{\vk}(\eta_0,\eta_0) = B_{\vk}(\eta_0,\eta_0)=0$ and 
$A_{\vk}(\eta_0,\eta_0) = D_{\vk}(\eta_0,\eta_0)=1$.
Clearly, each of the Hamiltonians (\ref{H1})--(\ref{H3}) 
leads to different $f_{k}(\eta)$ and $g_{k}(\eta)$ all satisfying (\ref{wr1}). 

The time $\eta_0(k)$ will be on the NPH (see below) and different Hamiltonians will
be minimized at that time. It should already be clear at
this point that the two-point function of, say, $\hat{\psi}_{\vk}$  will depend on the choice
of the initial state through
Eqs. (\ref{trans}) and (\ref{abcd}), since minimizing 
different Hamiltonians corresponds to imposing  different conditions on the  
field operators at $\eta_0(k)$.

In order to deal with simple mode functions, we will confine our attention to the case of an 
inflationary background with a  scale factor  parametrized, in conformal time, as 
 \begin{equation}
a(\eta) = \biggl( -  \frac{\eta}{\eta_1}\biggr)^{-\beta}, ~~~~\eta<-\eta_{1},
\label{dS}
\end{equation}
where $\eta_{1}$ marks the end of the inflationary epoch.
The pure de Sitter case corresponds to 
 $\beta = 1$. The inequality $k/a(\eta) \leq \Lambda $ is 
saturated, by definition, at the time $\eta_0(k)$. We shall refer to this time as the time of exit
from the NPH, not to be confused, of course, 
with the more standard ``horizon-exit" time $\eta_{\rm ex}$. In our case,
\begin{equation}
\eta_{0}(k) = - \eta_{1} \biggl(\frac{\Lambda}{k}\biggr)^{1/\beta}.
\end{equation}

Let us then discuss the minimization of the different Hamiltonians starting with (\ref{H1})
 but suppressing the label $^{(1)}$ for simplicity.
In Fourier space 
\begin{equation}
\hat{H}_{k}(\eta) = \frac{1}{4} \biggl[\frac{1}{a^2}(\hat{\Pi}_{\vk} \hat{\Pi}^{\dagger}_{\vk} + 
\hat{\Pi}_{\vk}^{\dagger} \hat{\Pi}_{\vk}) + k^2 a^2(\hat{\Psi}_{\vk} \hat{\Psi}^{\dagger}_{\vk} + 
\hat{\Psi}_{\vk}^{\dagger} \hat{\Psi}_{\vk}) \biggr],
\label{H1k}
\end{equation}
with $\hat{H}(\eta) = \int d^3 k  \hat{H}_{k}(\eta)$.  
The appropriately normalized mode functions are
\begin{eqnarray}
f_{k}(\eta) &=& \frac{{\cal N}}{a(\eta)\sqrt{2 k}} \sqrt{-x} H^{(1)}_{\mu}(- x) ~ , ~~ 
g_{k}(\eta) =  a^2(\eta) f_{k}',
\label{fk}\\
{\cal N} &=& \sqrt{\frac{\pi}{2}} e^{\frac{i}{2}(\mu + 1/2)\pi},~~~~~\beta = \mu -\frac{1}{2},
\label{N}
\end{eqnarray}
where $ x = k \eta$,  
and $H_{\mu}^{(1)}( -x)$ are the first-order Hankel functions of index $\mu$ \cite{magn}.
In order to minimize the Hamiltonian (\ref{H1}) let us consider the auxiliary operator 
\begin{equation}
\hat{Q}_{\vk} = \frac{1}{\sqrt{2 k}} \biggl[\frac{\hat{\Pi}_{\vk}}{a} -
i a k \hat{\Psi}_{\vk} \biggr].
\label{Q1}
\end{equation}
Equation (\ref{Q1})
allows Eq. (\ref{H1k})   to be expressed as 
\begin{equation}
\hat{H}_{k} = \frac{k}{2}\biggl[ \hat{Q}^{\dagger}_{\vk} \hat{Q}_{\vk} 
+ \hat{Q}_{\vk} \hat{Q}^{\dagger}_{\vk}\biggr],
\label{min1}
\end{equation}
while
canonical commutation relations between
conjugate field operators,
\begin{equation}
[ \hat{\psi}(\vec{x},\eta), \hat{\pi}(\vec{y},\eta)] = 
i \delta^{(3)} (\vec{x} - \vec{y}),
\label{cancomm}
\end{equation}
imply
 $[\hat{Q}_{\vk}, \hat{Q}_{\vp}^{\dagger} ] = \delta^{(3)}(\vec{k} - \vec{p})$.
Consequently, the state minimizing (\ref{H1}) is the one annihilated by $\hat{Q}_{\vk}$ (provided
it is normalizable).

In order to evaluate the corrections induced on the two-point function by 
this particular initial state, we have to compute 
\begin{equation}
\langle 0^{(1)},\eta_0| \hat{h}(\vec{x},\eta) \hat{h}(\vec{y},\eta) | \eta_{0},0^{(1)}\rangle \equiv 
\frac{ \ell_{\rm P}^2}{4\pi^3} \int d^3 k~\int d^3 p 
\langle ~\hat{\Psi}_{\vk}(\eta)~ \hat{\Psi}_{\vp}(\eta)~ \rangle e^{ - i (\vk\cdot\vx + \vp\cdot\vy)},
\label{twopo1}
\end{equation}
where $\langle...\rangle \equiv\langle0^{(1)},\eta_0|...| \eta_{0},0^{(1)}\rangle $ 
means that the expectation values should be evaluated 
over the state minimizing $H^{(1)}$ at the time $\eta_0$. Inserting Eqs. (\ref{trans}) into Eq. (\ref{twopo1}),
we obtain
\begin{eqnarray}
&&\langle 0^{(1)},\eta_0| \hat{h}(\vec{x},\eta) \hat{h}(\vec{y},\eta) | \eta_{0},0^{(1)}\rangle =
\frac{ \ell_{\rm P}^2}{4\pi^3}  \int d^{3} k \int d^{3} p \biggl[ A_{k}(\eta,\eta_0) A_{p}(\eta,\eta_0) 
\langle\hat{\Psi}_{\vk}(\eta_0) \hat{\Psi}_{\vp}(\eta_0)\rangle 
\nonumber\\
&&+ 
 B_{k}(\eta,\eta_0) B_{p}(\eta,\eta_0) 
\langle\hat{\Pi}_{\vk}(\eta_0) \hat{\Pi}_{\vp}(\eta_0)\rangle + 
 B_{k}(\eta,\eta_0) A_{p}(\eta,\eta_0) 
\langle\hat{\Pi}_{\vk}(\eta_0) \hat{\Psi}_{\vp}(\eta_0)\rangle +
\nonumber\\
&&+  A_{k}(\eta,\eta_0) B_{p}(\eta,\eta_0) 
\langle\hat{\Psi}_{\vk}(\eta_0) \hat{\Pi}_{\vp}(\eta_0)\rangle \biggr]e^{ - i (\vk\cdot\vx + \vp\cdot\vy)}.
\label{intermediate}
\end{eqnarray}

The various expectation values appearing in (\ref{intermediate}) can be computed using 
the relation between the canonical operators (evaluated at the initial time $\eta_0$) 
and the operators (\ref{Q1}) annihilating the initial state. 
Defining now the power spectrum (i.e.  
the Fourier transform of the two-point function) by, 
\begin{equation}
\langle 0^{(1)},\eta_0| \hat{h}(\vec{x},\eta) \hat{h}(\vec{y},\eta) |\eta_{0}, 0^{(1)}\rangle 
= \int \frac{d k}{k} |\delta_{h}(k,\eta)|^2 
\frac{\sin{k r} }{k r},
\label{ps1}
\end{equation}
we obtain, from Eq. (\ref{intermediate}) and 
with the help of Eqs. (\ref{fk})--(\ref{N}),
\begin{equation}
|\delta_{h}(k, \eta)|^2 = \frac{2^{4 \beta -1}}{ \pi^3  } (  2 \beta )^{ - 2 \beta}
\Gamma(\beta + 1/2)^2 
\biggl(\frac{H_1}{M_{\rm P}}\biggr)^2   
\biggl(\frac{k}{k_1}\biggr)^{2(1 - \beta)} \biggl[ 1  +
\frac{\beta}{ x_{0} } \sin {(2 x_0 + \beta \pi)}\biggr],
\label{psd}
\end{equation}
where we denote by $H_1$ the Hubble parameter at the end of inflation. Finally, 
$\frac{k}{k_1} = \frac{\omega}{\omega_1}$ is the ratio of the generic proper frequency
to the one corresponding to the end-point of the spectrum $\omega_1 = H_1/a$.

In order to derive Eq. (\ref{psd}) the limit $ x =k \eta \ll 1$, corresponding to looking
 at the correlation function at late times, has been taken. 
Also, since $|x_0| \gg 1$, only the leading correction in $1/x_0$ has been kept.
 Furthermore, using  $k/a(\eta_0) = \Lambda$, according to Eq. 
(\ref{dS}) the initial 
time $\eta_0$ can be easily related to the value of $\Lambda$ by
\begin{equation}
|x_0| = |k\eta_{0}| =  \beta \frac{\Lambda}{H_{\rm ex}^{\rm NPH}},
\label{NPH}
\end{equation}
where $H_{\rm ex}^{\rm NPH} = H(t_0(k))$ denotes the Hubble parameter 
at the time $t_0(k)$ when a given scale ``exits'' the NPH.  
Note that $x_0$ depends on $k$ except in the case 
of pure de Sitter.
We see that, as a consequence,   corrections
 to the standard results are larger at small $k$ for power law inflation 
(corresponding to larger values of $H_{\rm ex}^{\rm NPH}$), while the opposite is true for
superinflation ($0 < \beta < 1$).
Equation (\ref{psd}) generalizes the result obtained in \cite{dan} in the case of pure 
de Sitter space. Indeed, for $\beta = 1$, Eq. (\ref{psd}) gives 
exactly 
\begin{equation}
|\delta_{h}(k,\eta)|^2 = \frac{1}{2\pi^2} \biggl(\frac{H}{M_{\rm P}}\biggr)^2 
\biggl[ 1 - \frac{\sin{2 x_0}}{x_0}\biggr].
\end{equation}

Let us now repeat the same procedure in the case of  $H^{(2)}$, which is, incidentally,
 the Hamiltonian used 
in \cite{dan}. In the case of (\ref{H2}) we have 
\begin{equation}
\hat{H}_{k}(\eta) = \frac{1}{4} \biggl[(\hat{\pi}_{\vk} \hat{\pi}^{\dagger}_{\vk} + 
\hat{\pi}_{\vk}^{\dagger} \hat{\pi}_{\vk}) + k^2 (\hat{\psi}_{\vk} \hat{\psi}^{\dagger}_{\vk} + 
\hat{\psi}_{\vk}^{\dagger} \hat{\psi}_{\vk}) + k F(x)( \hat{\pi}_{\vk} \hat{\psi}_{\vk}^{\dagger}+ 
\hat{\pi}_{\vk}^{\dagger} \hat{\psi}_{\vk} +  \hat{\psi}_{\vk} \hat{\pi}_{\vk}^{\dagger}+ 
\hat{\psi}_{\vk}^{\dagger} \hat{\pi}_{\vk})\biggr],
\label{ham3}
\end{equation} 
where
\begin{equation}
k F(x) = {\cal H}.
\label{pump}
\end{equation}
Solving the evolution in the Heisenberg picture,  the mode functions can be written as \cite{magn}
\begin{eqnarray}
f_{k}(\eta) &=& \frac{{\cal N}}{\sqrt{2 k}} \sqrt{-x} H^{(1)}_{\mu}(- x), 
\label{fk1}\\
g_{k}(\eta) &=&  - {\cal N}\sqrt{\frac{k}{2}} \sqrt{-x} H^{(1)}_{\mu -1} (-x).
\label{gk1} 
\end{eqnarray}
In order to minimize the Hamiltonian (\ref{ham3}) at the initial time $\eta_{0}$, 
we introduce
\begin{equation}
\hat{Q}_{\vk} = \frac{1}{\sqrt{2 k}} \biggl[e^{ -i \gamma } \hat{\pi}_{\vk} -
i e^{  i \gamma } k \hat{\psi}_{\vk} \biggr],
\label{Q}
\end{equation}
where $\gamma $ is a time-dependent parameter. 
 Using Eq. (\ref{Q}), the Hamiltonian (\ref{ham3}) 
can be put in the same form as (\ref{min1}) provided the following relation is imposed
 between $\gamma $ and $F(x)$ of Eq. (\ref{ham3}): 
\begin{equation}
\sin{2 \gamma } = F(x).
\label{alphatoF}
\end{equation} 
The canonical commutation relations Eq. (\ref{cancomm}) now imply  
$[\hat{Q}_{\vk}, \hat{Q}_{\vp}^{\dagger}] = 
\cos{2\gamma } \delta^{(3)}(\vec{k} -\vec{p})$,
so that the initial state  minimizing 
(\ref{ham3}) is again the one annihilated by $\hat{Q}_{\vk}$.

The wave-functional 
of the initial state can be easily derived and, for each mode, it has a Gaussian form:
\begin{equation}
\Phi [ \psi_{\vk}] = N {\rm exp}
 \left( - \sum_k \frac{k}{2} (\psi_{\vk} \psi_{-\vk}) e^{-2i\gamma} \right).
\label{wf}
\end{equation}
This state is  normalizable provided $|\gamma |< \pi/4$. 
Using Eq. (\ref{alphatoF}), we see that $|\gamma | = \pi/4$
corresponds to a time $\eta_0$ for which $|F(x_0)| =1$, which is basically equivalent 
to the condition of (standard) horizon crossing. Consequently,  provided the modes of the field 
are inside the horizon at the ``initial" time $\eta_0$, the state (\ref{wf}) is normalizable.

The two-point function to be computed  now is 
\begin{equation}
\langle 0^{(2)},\eta_0| \hat{h}(\vec{x},\eta) \hat{h}(\vec{y},\eta) | \eta_{0},0^{(2)}\rangle =
 \frac{ \ell_{\rm P}^2}{4 \pi^3~a(\eta)^2} \int d^3 k~\int d^3 p 
\langle ~\hat{\psi}_{\vk}(\eta)~ \hat{\psi}_{\vp}(\eta)~ \rangle e^{ - i (\vk\cdot\vx + \vp\cdot\vy)}
,\label{twopo}
\end{equation}
and the related power spectrum evaluated at late times ($ x =k \eta \ll 1$) is
\begin{equation}
|\delta_{h}(k, \eta)|^2 =  \frac{2^{4 \beta -1}}{ \pi^3  } ( 2 \beta )^{ - 2 \beta}
\Gamma(\beta + 1/2)^2 
\biggl(\frac{H_1}{M_{\rm P}}\biggr)^2   
\biggl(\frac{k}{k_1}\biggr)^{2(1 - \beta)} 
\biggl[ 1 - \beta \frac{\cos{(2 x_0 + \beta \pi)}}{2 x_0^2}\biggr].
\label{psus1}
\end{equation}
In the  de Sitter case, $\beta = 1$, Eq. (\ref{psus1}) gives 
\begin{equation}
|\delta_{h}(k, \eta)|^2 = \frac{1}{2 \pi^2} \biggl(\frac{H}{M_{P}}\biggr)^2
\biggl[ 1 + \frac{\cos{2 x_0}}{2 x_0^2}\biggr].
\label{finds}
\end{equation}

Note that, in \cite{dan}, use was made of the
 Hamiltonian  $H^{(2)}$, but the initial state was such as to minimize $H^{(1)}$. Thus, in agreement with
our general statements, the results found in  \cite{dan}, were the same (in de Sitter) as those we found
unsing $H^{(1)}$ and minimizing it. What matters is not which Hamiltonian is used for the evolution, but
which one is used for defining the initial state through an energy minimization procedure!

The Hamiltonian (\ref{H3}) can be minimized following the same procedure 
already discussed in the case of Eqs. (\ref{H1}) and (\ref{H2}). 
Defining the function 
\begin{eqnarray}
\omega^2(x) = \biggl( 1 - \frac{1}{k^2} \frac{a''}{a}\biggr),
\end{eqnarray}
the Hamiltonian (\ref{H3}) can be written in the simple form \footnote{From now on the tilde in the momentum operators 
will be omitted for the sake of simplicity.}
\begin{equation}
\hat{H}_{k}(\eta) = \frac{1}{4} \biggl[(\hat{\pi}_{\vk} \hat{\pi}^{\dagger}_{\vk} + 
\hat{\pi}_{\vk}^{\dagger} \hat{\pi}_{\vk}) + k^2\omega^2(x) (\hat{\psi}_{\vk} \hat{\psi}^{\dagger}_{\vk} + 
\hat{\psi}_{\vk}^{\dagger} \hat{\psi}_{\vk})\biggr].
\label{ham4}
\end{equation}
Defining now the  operator
\begin{equation}
\hat{Q}_{\vk} = \frac{1}{\sqrt{2 k}}\biggl[ \hat{\pi}_{\vk} - i k  \omega \hat{\psi}_{\vk}\biggr],
\label{Qpr}
\end{equation}
the Hamiltonian can  again be expressed in the same  form 
previously discussed, namely, the one given by Eq. (\ref{min1}) with the caveat that 
now the operator (\ref{Qpr}), if compared to the one defined in Eq. (\ref{Q})
 has a different expression in terms of the 
canonical fields.
The commutation relations are now
 $[Q_{\vk}, Q_{\vp}^{\dagger}] = \omega \delta^{(3)}(\vec{k}-\vec{p})$.
The mode functions
  $f_{k}(\eta)$ are the same as the ones given in Eq.(\ref{fk1}), while  ${g}_{k}$ is given by
\begin{eqnarray}
g_{k}(\eta) &=&  - {\cal N}\sqrt{\frac{k}{2}} \sqrt{-x}\biggl[ H^{(1)}_{\mu -1} (-x) +
\frac{(1 -2 \mu)}{2(-x)} H^{(1)}_{\mu} (-x)\biggr] ,
\label{tgk} 
\end{eqnarray}
Repeating the steps used in the two previous cases we arrive at the power spectrum:
\begin{equation}
|\delta_{h}(k,\eta)|^2 = \frac{2^{4 \beta -1}}{ \pi^3  } ( 2 \beta )^{ - 2 \beta}
\Gamma(\beta + 1/2)^2 
\biggl(\frac{H_1}{M_{\rm P}}\biggr)^2   
\biggl(\frac{k}{k_1}\biggr)^{2(1 - \beta)}\biggl[ 1 - 
\frac{  \beta (\beta+1)}{2 x_0^3} \sin{( 2 x_0 + \pi\beta)}
\biggr].
\label{ps3}
\end{equation}
In the de Sitter case, we have:  
\begin{equation}
|\delta_{h}(k,\eta)|^2 = \frac{1}{2 \pi^2} \biggl(\frac{H}{M_{\rm P}}\biggr)^2 
\biggl[ 1 + \frac{\sin{2 x_0}}{x_0^3}\biggr].
\label{psd3a}
\end{equation}
Comparing Eq. (\ref{ps3}) with 
(\ref{psus1}) we see that the corrections are even smaller than the ones obtained using the Hamiltonian of Eq. (\ref{ham3}).
Furthermore, both Eqs. (\ref{ps3}) and (\ref{psus1}) lead to effects smaller than (\ref{psd}). 

We have shown that transplanckian effects can be tamed by choosing a sufficiently adiabatic Hamiltonian.
We may ask whether the opposite can be achieved, i.e. enhance transplanckian effects
 by minimizing
particularly contrived Hamiltonians  on the NPH. The answer turns out to be positive. Consider the 
following generating functional 
\begin{equation}
{\cal F}_{3\to 4}(\psi,\tilde{\Pi}) = \int d^{3} x\biggl[\psi \tilde{\Pi} - \frac{\tilde{\Pi}^2}{2 {\cal H}}\biggr].
\end{equation}
This rather weird canonical transformation allows the passage from $H^{(3)}$ to an $H^{(4)}$, where the new 
fields $\tilde{\Psi}$ and the new momenta $\tilde{\Pi}$ are 
\begin{equation}
\tilde{\Psi} = \psi - \frac{\tilde{\Pi}}{\cal H},~~~~\tilde{\Pi} = \tilde{\pi}.
\end{equation}
Without going through the details of the derivation, the following highly ``transplanckian''
Hamiltonian can be obtained:
\begin{equation}
H^{(4)} = \int d^3 x \frac{1}{2} \biggl[ \frac{k^2}{{\cal H}^2} \tilde{\Pi}^2 + 2 \biggl( \frac{k^2}{\cal H} - {\cal H} - 
\frac{{\cal H}'}{{\cal H}} \biggr) \tilde{\Pi} \tilde{\Psi}+ \biggl(k^2 - \frac{a''}{a}\biggr) \tilde{\Psi}^2 \biggr].
\end{equation}
The same procedure as described for the other three Hamiltonians can be repeated, and the correction
induced in the power spectrum turns out to be of order $1$ in $1/x_0$, namely (here we just give the result
in the de Sitter case):
\begin{equation}
|\delta_{h}(k,\eta)|^2 = \frac{1}{4\sqrt{2}~\pi^2} \biggl(\frac{H}{M_{\rm P}}\biggr)^2[ 3 - \cos{2 x_0}].
\end{equation} 

The question thus arises 
 of which  Hamiltonian, if any, should be minimized on the NPH. 
  A similar problem has been discussed in the past \cite{weiss,fulling}, 
though in a somewhat different context. The idea was to study the properties of two-point functions
at arbitrary space-time points in the state $|0, t_0 \rangle$, minimizing a certain
 Hamiltonian at a given (mode-independent) ``initial time" time $t_0 $, i.e.
\begin{equation}
\langle  t_0,0| 
 \hat{h}(x)\hat{h}(y)|0, t_0 \rangle \; ,
\label{exv}
\end{equation}
where $x\equiv(\vec{x},\eta)$ and $y \equiv(\vec{y},\eta')$.
 It was found \cite{weiss}  that, only if the state $|0, t_0 \rangle$ minimizes a
 very special class of Hamiltonians (in our case
just $H^{(3)}$), do the  singularities in (\ref{exv}) lie on the light cone  $(x-y)^2 =0$ and are 
of the Hadamard form.
All other Hamiltonians lead to additional  singularities outside the light cone.

It is easy to generalize our discussion to 
two-point functions  at unequal times.
 However, the situation 
discussed in \cite{weiss} is quite different from the one described here. On the one
hand we do impose a momentum cut-off. By itself, this would only smooth out the
singularity without really removing the phenomenon. A more significant
distinction is that we minimize different pieces of the Hamiltonian at different times. 
Direct calculations show that, unfortunately, 
 Hamiltonians (\ref{H2}) and (\ref{H3}) 
cannot be distinguished by using the singularity arguments of \cite{weiss}.

 There is probably a more physical reason why (\ref{ham4}) should be 
preferred over (\ref{ham3}). In  the case of (\ref{ham3}) the ``interacting'' part of the Hamiltonian 
vanishes as $1/\eta$ for large $\eta$. In other words, the switching-on of the interaction
at finite $\eta$ is faster 
for $H^{(2)}$ than it is for $H^{(3)}$. The latter is definitely a more ``adiabatic" Hamiltonian.
It thus looks reasonable to assume that the truly initial state remains in the ground state of 
this ``very adiabatic" Hamiltonian.

Our conclusions can be summarized as follows:
\begin{itemize}
\item In case where one has serious doubts about the theory of cosmological perturbations at
$t \to - \infty$ it looks reasonable to assign ``initial" quantum fluctuation on a NPH 
to the future of which the standard procedure applies.
\item Although there are ambiguities in defining the Hamiltonian in 
time-dependent problems, physical results do not depend on the choice of such Hamiltonian.
 Corrections to the standard result only depend on the choice of the `` initial" state on the NPH.
\item An attractive, though by no means unique, choice for the initial state is the one
 that minimized the Hamiltonian on the NPH. The so-defined state {\em does depend} on how the Hamiltonian
to be minimized is chosen.
\item The size of transplanckian corrections is a sensitive function of 
that choice with the most  ``adiabatic" Hamiltonians giving the smallest corrections in terms of the
small parameter $H_{\rm ex}^{\rm NPH} / \Lambda$. The  most adiabatic Hamiltonian, 
whose minimization on a space-like hypersurface leads to correlation functions of the Hadamard form,
gives transplanckian corrections that are much smaller than previously suggested.
\item A final answer to the question of transplanckian corrections will only be found 
when a theory of transplanckian physics is defined, \dots and solved.
\end{itemize}

It is a pleasure to thank M. Gasperini for useful discussions and correspondence. V.B. would 
like to thank the Theory Division of CERN for hospitality while this work was 
carried out.

\end{document}